\documentclass[aps,preprint,nofootinbib,superscriptaddress,prc]{revtex4}
\usepackage{amssymb}
\usepackage{CJK}
\usepackage{indentfirst}
\usepackage{amsmath}
\usepackage{xcolor}
\usepackage{epsfig}
\usepackage[bookmarksnumbered,bookmarksopen,colorlinks,citecolor=blue,linkcolor=blue] {hyperref}
\begin{document}

\date{\today}

\title{Enhancement of $\alpha$-decay by positive hexadecapole deformation}
\author{Kai Ren}
\affiliation{Guangxi Key
Laboratory of Nuclear Physics and Technology,  Guilin 541004, People's Republic of China}
\affiliation{Department of Physics, Guangxi Normal University, Guilin 541004, People's Republic of China}

\author{Minghui Hu}
\affiliation{Guangxi Key
Laboratory of Nuclear Physics and Technology,  Guilin 541004, People's Republic of China}
\affiliation{Department of Physics, Guangxi Normal University, Guilin 541004, People's Republic of China}

\author{Pengfei Ma}
\affiliation{Guangxi Key
Laboratory of Nuclear Physics and Technology,  Guilin 541004, People's Republic of China}
\affiliation{Department of Physics, Guangxi Normal University, Guilin 541004, People's Republic of China}

\author{Junlong Tian}
 \email{tianjl@gxnu.edu.cn}
 \affiliation{Guangxi Key
Laboratory of Nuclear Physics and Technology,  Guilin 541004, People's Republic of China}
\affiliation{Department of Physics, Guangxi Normal University, Guilin 541004, People's Republic of China}

\author{Cheng Li}
 \email{licheng@gxnu.edu.cn}
\affiliation{Guangxi Key
Laboratory of Nuclear Physics and Technology,  Guilin 541004, People's Republic of China}
\affiliation{Department of Physics, Guangxi Normal University, Guilin 541004, People's Republic of China}

\begin{abstract}
Whether hexadecapole deformation ($\beta_4$) influences $\alpha$ decay remains 
controversial: machine-learning analyses suggest a strong link to cluster 
preformation, while empirical formulas find only marginal effects. 
We show that this discrepancy originates in the treatment of shell effects---without an explicit 
shell correction, residuals near magic numbers are absorbed into the 
deformation coefficients, obscuring the genuine $\beta_4$ dependence. 
Adding the inverse Casten factor $C_{pn}$, which encodes valence proton--neutron 
correlations relative to the nearest closed shells, and the parent-nucleus deformation 
$\beta_4^{(p)}$ to the Royer formula reduces 
the root-mean-square deviation from 0.309 to 0.184 for 192 even--even nuclei. 
The fitted negative $\beta_4^{(p)}$ coefficient shows that positive 
hexadecapole deformation systematically shortens half-lives, consistent 
with enhanced $\alpha$-cluster preformation at locally convex surface 
regions. For the $Z=94$ isotopic chain (Pu), where pronounced $\beta_4^{(p)}>0$ 
occurs, the original Royer formula overestimates half-lives by up to 
$\sim\!0.5$~dex, providing a clear, testable signature of this 
surface-preformation effect. The same correction also improves the UDL and 
yields predictions for 1060 even--even nuclei.
\end{abstract}

\pacs{23.60.+e, 03.65.Xp, 21.10.Tg, 21.60.Gx}
 \keywords{nuclear potential depth, $\alpha$-decay, shell correction energy}%

\maketitle

\section{Introduction}
$\alpha$ decay provides a sensitive probe of nuclear structure, since its
half-life is governed jointly by $\alpha$-cluster preformation and
Coulomb-barrier penetration~\cite{Qi2019,Andreyev2013,Qi2009Micro,NiRen2010,Mirea2020}; it is also the primary
tool for identifying the decay chains of superheavy nuclei. Since the
Geiger--Nuttall law~\cite{Geiger1911}, a variety of phenomenological formulas
have been proposed, including the Viola--Seaborg relation~\cite{Viola1966}, the
Royer formula~\cite{Royer2000}, the universal decay law~\cite{Qi2009}, and the
unified formula of Ni \textit{et al.}~\cite{Ni2008}; the Royer formula was later
generalized to a unitary form~\cite{Deng2020,Ren2026}. While these expressions capture
the dominant $Q_\alpha$ dependence, systematic residuals persist near shell
closures and for strongly deformed nuclei, indicating that additional
structural information is required\cite{Qi2010Collective,XuRen2007,IsmailAdel2022}.
Among these global parameterizations, the Royer formula is adopted here as the 
primary framework because it offers an optimal balance between analytical 
transparency and accuracy.

Nuclear deformation is one such ingredient. Quadrupole deformation $\beta_2$
lowers the Coulomb barrier and thereby shortens half-lives, an effect already
incorporated into several empirical formulas~\cite{XuRen2006,Denisov2024,Ismail2025,You2025,Dahmardeh2017,Delion2015,NiRen2015,You2024,XuRen2006DDCM,IsmailAdel2014}.
Hexadecapole deformation $\beta_4$, by contrast, has received far less
attention: because it is typically an order of magnitude smaller than $\beta_2$,
it has usually been treated as a higher-order correction in the multipole
expansion, and its independent role has rarely been examined. Recent work
disagrees on how important that role is. Machine-learning
analyses~\cite{Ma2026} report strong correlations between $\beta_4$ and both the
decay energy $Q_\alpha$ and the preformation probability $P_\alpha$, and
interpret them as evidence that nuclei with hexadecapole deformation are more
likely to form $\alpha$ clusters, with the effect propagating through
parent--daughter shape inheritance. Ismail \textit{et al.}~\cite{Ismail2025}, on
the other hand, find that adding $\beta_4$ to five empirical formulas reduces
the RMSD by less than $0.7\%$, a gain they regard as marginal. Whether
$\beta_4$ carries genuine physical significance for $\alpha$ decay, or merely
acts as a negligible correction, thus remains an open question.

\begin{figure}[htp]
	\centering
	\includegraphics[width=1.0\linewidth]{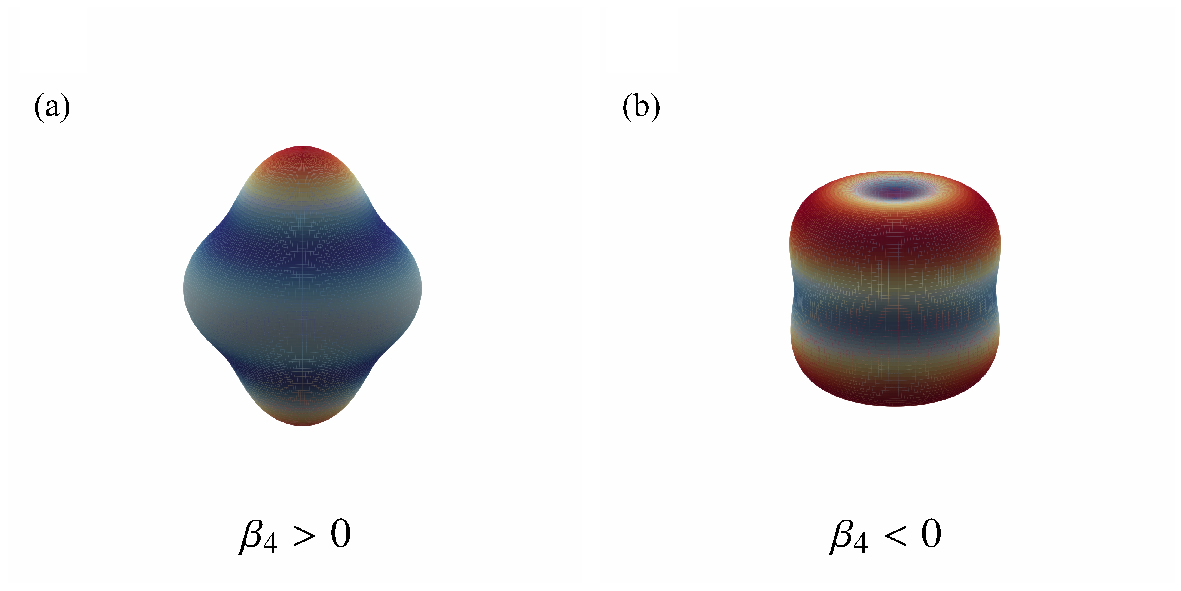}
	\caption{(Color online) Schematic nuclear shapes for (a) positive and (b) negative hexadecapole deformation. Positive $\beta_4$ produces locally convex surface regions, whereas negative $\beta_4$ produces relatively concave regions.}
	\label{fig:shape}
\end{figure}

We argue that this discrepancy originates mainly in the treatment of shell
effects. When an explicit shell correction is absent, the fitted $\beta_4$ coefficient inevitably absorbs part of the shell-induced residuals, especially near magic numbers; consequently, the extracted deformation dependence becomes obscured, even if the overall fit quality appears improved. 
Thus, isolating the shell effect is a prerequisite for uncovering the genuine role of $\beta_4$. 
Yet neither the original Royer formula~\cite{Royer2000} nor the formulas used in Ref.~\cite{Ismail2025} include such a correction, so the prominent residuals around $N=126$ are partially reabsorbed into the deformation coefficients, biasing the extracted $\beta_4$ dependence. 
A second issue is the choice of deformation partner: Ref.~\cite{Ismail2025} employs the daughter deformation $\beta_4^{(d)}$, 
whereas $\alpha$-cluster 
preformation is a process localized on the parent-nuclear surface, which makes 
the parent deformation $\beta_4^{(p)}$ the physically appropriate input. 
Indeed, a positive $\beta_4^{(p)}$ creates locally convex surface regions 
(Fig.~\ref{fig:shape}) that are expected to favour four-nucleon correlations---a 
higher-order effect qualitatively different from the global elongation described 
by $\beta_2$.

To disentangle the shell and deformation effects identified above, we extend the Royer formula by including both the inverse Casten shell-correction 
factor $C_{pn}$ and the parent-nucleus hexadecapole deformation $\beta_4^{(p)}$. 
Fitting 192 even--even $\alpha$-decay
half-lives confirms that part of the shell residual is otherwise absorbed by
$\beta_4$: once $C_{pn}$ is included, the $\beta_4$ coefficient decreases from
$6.712$ to $5.986$ (by $\simeq 11\%$), while its sign and magnitude remain
stable. With the shell correction in place, $\beta_4^{(p)}$ shows a clean,
sign-dependent effect: positive $\beta_4$ systematically shortens half-lives,
consistent with enhanced cluster preformation, whereas negative $\beta_4$ acts
in the opposite direction. This behavior agrees with the machine-learning
findings of Ref.~\cite{Ma2026}. The resulting formula lowers the RMSD for
even--even nuclei from $0.309$ (original Royer) to $0.184$, and is further used
to predict the half-lives of 1060 even--even nuclei with the hexadecapole deformation.
\section{\label{sec:level2}Theoretical framework}
\subsection{Royer formula and its limitations}

The original Royer formula~\cite{Royer2000} reads
\begin{equation}
\log_{10}T_{1/2}^{\rm Royer} = a_0 + b_0 A^{1/6}\sqrt{Z}
 + c_0\frac{Z}{\sqrt{Q_\alpha}},
\label{eq:royer}
\end{equation}
where $A$, $Z$, and $Q_\alpha$ denote the mass number, proton number, and decay
energy of the parent nucleus, and the coefficients for even--even nuclei are
$a_0=-25.31$, $b_0=-1.1629$, and $c_0=1.5864$. Physically,
Eq.~(\ref{eq:royer}) is rooted in the Gamow picture of barrier
penetration~\cite{Gamow1928}: the $Z/\sqrt{Q_\alpha}$ term reproduces the
tunneling factor, while the $A^{1/6}\sqrt{Z}$ term effectively parameterizes the
structural dependence, including the barrier geometry and the average
$\alpha$-preformation probability. The formula carries no explicit dependence on
shell effects or nuclear deformation, however, and consequently leaves
systematic residuals in specific regions of the nuclear
chart~\cite{Denisov2024,Ismail2025,You2025,Tian2024Shell,Ren2026}.

\subsection{Improved formula with shell correction and parent-nucleus $\beta_4$}

To remedy these two deficiencies, we propose the improved formula
\begin{equation}
\log_{10}T_{1/2} = a + b A^{1/6}\sqrt{Z} + c\frac{Z}{\sqrt{Q_\alpha}}
 + \left(d\,\beta_4^{(p)} + C_{pn}\right),
\label{eq:improved}
\end{equation}
in which the two additional terms are defined as follows.

(i) \emph{Inverse Casten shell-correction factor $C_{pn}$.} This term
accounts for the valence proton--neutron correlations near shell
closures~\cite{Casten1985}:
\begin{equation}
C_{pn} = \frac{1}{2\max(N_p,1)} + \frac{1}{2\max(N_n,1)},
\label{eq:Cpn}
\end{equation}
where $N_p = \min(Z-Z_i,\,Z_{i+1}-Z)$ and
$N_n = \min(N-N_j,\,N_{j+1}-N)$ are the valence proton and neutron numbers
counted from the nearest closed shells ($Z_i = 28,\,50,\,82,\,126$;
$N_j = 28,\,50,\,82,\,126,\,184$)~\cite{Silisteanu2010Casten,Zhao2000NpNn},
and the regulator $\max(N_{p(n)},1)$ removes the divergence at shell
closures. Up to this regulator, Eq.~(\ref{eq:Cpn}) is the inverse of twice
the Casten factor $P = N_pN_n/(N_p+N_n)$, which motivates its name.
$C_{pn}$ is enhanced near closed shells, reflecting the reduced
collectivity and enhanced structural stability that suppress $\alpha$
decay. Compared with the shell-correction energy
$E_{\rm sh}=B_{\rm exp}-B_{\rm LD}$ adopted in
Refs.~\cite{Ren2026,Tian2024Shell}, $C_{pn}$ offers two advantages as a
shell indicator. First, $E_{\rm sh}$ inevitably contains deformation,
pairing, and other microscopic contributions in addition to the shell
effect, and thus cannot be interpreted strictly as a pure shell measure,
whereas $C_{pn}$ is constructed directly from the valence nucleon numbers
relative to the neighboring shell closures and therefore characterizes
shell-related structural variations more directly. Second, $C_{pn}$
depends only on $Z$, $N$, and the adopted magic numbers, without relying
on any specific nuclear mass model. Finally, the coefficient of $C_{pn}$
is fixed to unity, so that this term introduces no additional adjustable
parameter; when an independent coefficient is nevertheless allowed, the
fit returns a value close to unity, supporting the present choice.

(ii) \emph{Hexadecapole deformation term $d\,\beta_4^{(p)}$.} The parent-nucleus hexadecapole deformation $\beta_4^{(p)}$ is taken from
the WS4 mass model~\cite{Wang2014}. The essential difference from
Ref.~\cite{Ismail2025} is the use of the parent rather than the daughter
$\beta_4$, since $\alpha$-cluster preformation occurs on the parent
surface. It is also worth explaining why the quadrupole deformation
$\beta_2^{(p)}$ is not included in Eq.~(\ref{eq:improved}). The structural
term $A^{1/6}\sqrt{Z}$ in the Royer formula effectively parameterizes the
global nuclear size and the average barrier geometry; since quadrupole
deformation primarily modulates the global elongation and the average
Coulomb-barrier radius, its dominant effect is already absorbed, in a
statistical sense, by this term. Hexadecapole deformation, by contrast, is
a higher-order shape effect that this smooth term cannot capture: the
locally convex regions created by $\beta_4^{(p)}>0$ (Fig.~\ref{fig:shape})
enhance the nucleon--nucleon correlations at the nuclear surface and
thereby facilitate the assembly of four nucleons into an $\alpha$ cluster.
Consequently, $\beta_4^{(p)}$ carries structural information genuinely
distinct from the global quadrupole geometry, as will be confirmed by the
correlation analysis in Sec.~\ref{sec:results}.
\begin{figure}[htp]
	\centering
	\includegraphics[width=0.6\linewidth]{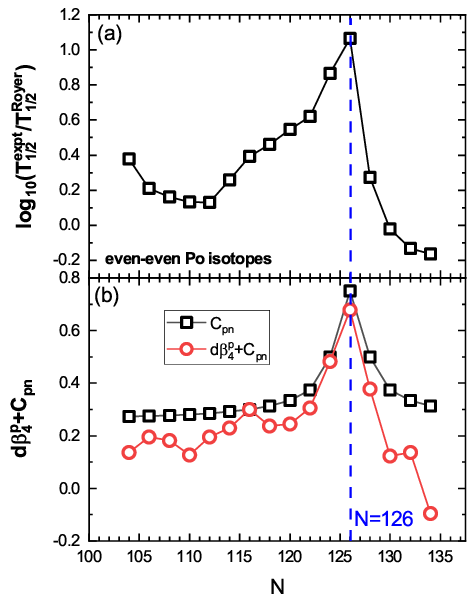}
	\caption{(Color online) Neutron-number dependence of the Royer residuals and the
	corresponding corrections for 16 even--even Po isotopes. Panel (a) shows
	$\log_{10}(T_{1/2}^{\rm expt}/T_{1/2}^{\rm Royer})$. Panel (b) shows the
	inverse Casten factor $C_{pn}$ and the combined correction
	$d\beta_{4}^{(p)}+C_{pn}$. The
	vertical dashed line marks the neutron shell closure at $N=126$.}
	\label{fig:Po_even}
\end{figure}

The parameters in Eq.~(\ref{eq:improved}) are determined by a least-squares fit
to the measured $\alpha$-decay half-lives of 192 even--even nuclei with
$52 \leq Z \leq 118$. We restrict the fit to
even--even nuclei because odd-$A$ and odd--odd nuclei involve additional
degrees of freedom---such as the unpaired-nucleon blocking effect and
nonzero angular momenta of the emitted $\alpha$ cluster---that would
introduce uncontrolled scatter and obscure the shell and hexadecapole
effects of interest here. In addition to the 175 entries taken from the NUBASE2020
evaluation~\cite{nubase2020}, the data set incorporates updated experimental data
for 10 nuclei, $^{108}$Xe~\cite{Auranen2018}, $^{146}$Sm~\cite{Chiera2024},
$^{154}$Dy~\cite{Chiera2022}, $^{174}$Hf~\cite{Belli2025},
$^{170}$Hg~\cite{Hilton2019}, $^{178}$Pb~\cite{Badran2016},
$^{216}$U~\cite{Zhang2021}, $^{218}$U~\cite{Zhang2021},
$^{286}$Fl~\cite{Gates2024}, and $^{288}$Fl~\cite{SamarkRoth2021}, together
with 7 newly measured nuclei, $^{160}$Os~\cite{Yang2024},
$^{214}$U~\cite{Zhang2021}, $^{272}$Hs~\cite{Oganessian2023},
$^{276}$Ds~\cite{Oganessian2023}, $^{284}$Cn~\cite{SamarkRoth2021},
$^{284}$Fl~\cite{Oganessian2025}, and $^{288}$Lv~\cite{Oganessian2025}. The
quality of the description is quantified by the root-mean-square deviation
(RMSD),
\begin{equation}
\sigma = \left[\frac{1}{n}\sum_{i=1}^{n}
 \left(\log_{10}T_{1/2,i}^{\rm exp} - \log_{10}T_{1/2,i}^{\rm calc}\right)^2
 \right]^{1/2}.
\label{eq:rms}
\end{equation}
The fit yields $a=-26.718\pm0.151$, $b=-1.138\pm0.006$, $c=1.608\pm0.003$, and
$d=-5.986\pm0.476$. The coefficient $d$ is negative and differs from zero by
more than ten standard deviations; a positive $\beta_4^{(p)}$ therefore shortens
the half-life, as anticipated above.

The interplay of the two new terms is illustrated in Fig.~\ref{fig:Po_even} for
the Po isotopic chain. The residuals of the original Royer formula peak at
$N=126$ and drop sharply beyond it. The combined correction, fitted to the same
16 Po isotopes, follows this neutron-number variation closely, whereas $C_{pn}$
alone---although it also peaks at $N=126$---fails to reproduce the residual
pattern away from the shell closure. The shell and hexadecapole-deformation
effects therefore have to be included simultaneously.

\subsection{Extension to other empirical formulas}

The correction $d\,\beta_4^{(p)}+C_{pn}$ is not tied to the functional form of
the Royer formula. As a representative example we consider the universal decay
law (UDL)~\cite{Qi2009},
\begin{equation}
\log_{10}T_{1/2}^{\rm UDL}
 = a_{\rm U}\,Z_{\alpha}Z_{d}\sqrt{\frac{\mu}{Q_\alpha}}
 + b_{\rm U}\sqrt{\mu Z_{\alpha}Z_{d}
   \left(A_{\alpha}^{1/3}+A_{d}^{1/3}\right)}
 + c_{\rm U},
\label{eq:UDL}
\end{equation}
where $Z_\alpha$ ($A_\alpha$) and $Z_d$ ($A_d$) are the charge (mass) numbers of
the $\alpha$ particle and the daughter nucleus, and
$\mu = A_\alpha A_d/(A_\alpha+A_d)$ is the reduced mass number. Appending the
same correction gives
\begin{equation}
\log_{10}T_{1/2}^{{\rm UDL}\beta}
 = \log_{10}T_{1/2}^{\rm UDL}
 + \left(d_{\rm U}\,\beta_4^{(p)} + C_{pn}\right).
\label{eq:UDLbeta}
\end{equation}
For a consistent comparison, all coefficients are refitted to the same set of
192 even--even nuclei, yielding $a_{\rm U}=0.411\pm0.002$,
$b_{\rm U}=-0.425\pm0.003$, and $c_{\rm U}=-21.798\pm0.254$ for
Eq.~(\ref{eq:UDL}), and $a_{\rm U}=0.416\pm0.001$, $b_{\rm U}=-0.418\pm0.002$,
$c_{\rm U}=-23.069\pm0.148$, and $d_{\rm U}=-6.889\pm0.470$ for
Eq.~(\ref{eq:UDLbeta}). The coefficient $d_{\rm U}$ is again negative and of the
same magnitude as $d$ in Eq.~(\ref{eq:improved}), indicating that the extracted
hexadecapole dependence is a property of the data rather than of a particular
parameterization. The same procedure was applied to the VSS~\cite{Viola1966}
and NRDX~\cite{Ni2008} formulas, with the RMSD decreasing in every case; the
results are summarized in Table~\ref{tab:formula_comparison} and discussed in
Sec.~\ref{sec:results}.

\section{Results and Discussion}
\label{sec:results}
\subsection{Selection of the correction terms}

\begin{figure}[htp]
	\centering
	\includegraphics[width=1\linewidth]{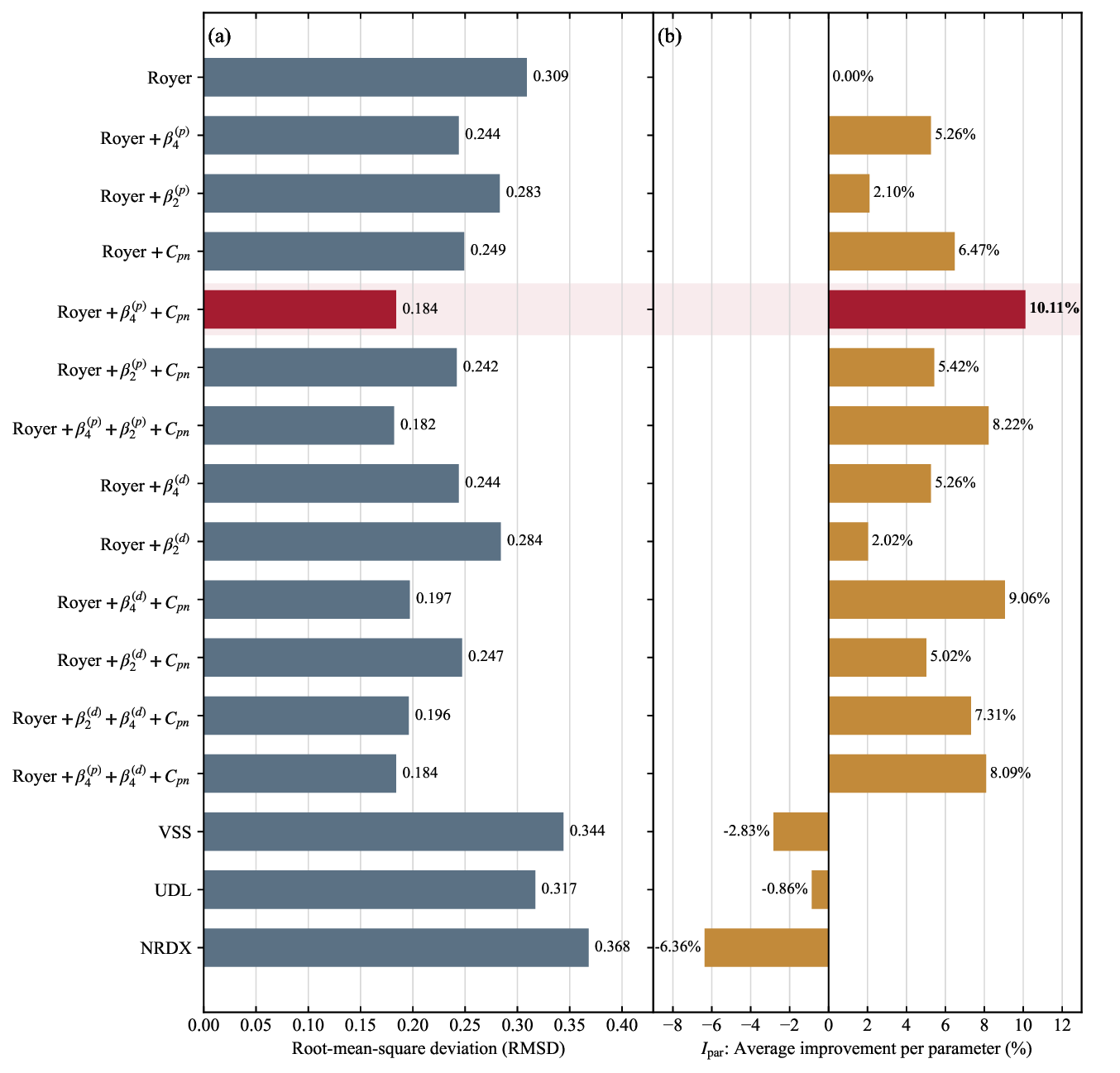}
	\caption{(Color online) Comparison of the fitting performance of the different formula
	variants: (a) RMSD and (b) $I_{\mathrm{par}}$ defined in
	Eq.~(\ref{eq:Ipar}). Red bars highlight the preferred formula,
	Eq.~(\ref{eq:improved}). The superscripts $(p)$ and $(d)$ denote the parent
	and daughter nuclei, respectively.}
	\label{fig:model_comparison}
\end{figure}

Figure~\ref{fig:model_comparison} compares the original Royer formula~\cite{Royer2000} with a
series of extensions containing the inverse Casten factor $C_{pn}$ and the
quadrupole or hexadecapole deformations of the parent and daughter nuclei,
denoted by the superscripts $(p)$ and $(d)$, respectively. Since the variants
differ in the number of free coefficients, we quantify the improvement per
fitting parameter by
\begin{equation}
I_{\mathrm{par}} =
\frac{(\sigma_{\rm Royer}-\sigma_{\rm Formula})/\sigma_{\rm Royer}}
{N_{\mathrm{par}}}\times 100\%,
\label{eq:Ipar}
\end{equation}
where $\sigma_{\rm Royer}=0.309$ is the RMSD of the original Royer formula, $\sigma_{\rm Formula}$ is the RMSD of
the corresponding extended formula, and $N_{\mathrm{par}}$ is the total number of fitting parameters of the formula
considered. Note that $C_{pn}$ enters with a fixed coefficient and therefore
does not increase $N_{\mathrm{par}}$.

Individually, both $C_{pn}$ and $\beta_4^{(p)}$ are effective: $C_{pn}$ reduces
the RMSD from $0.309$ to $0.249$ ($I_{\mathrm{par}}=6.47\%$), and
$\beta_4^{(p)}$ alone to $0.244$ ($5.26\%$), whereas $\beta_2^{(p)}$ achieves
only $0.283$ ($\approx\!2\%$).  Their combination yields $0.184$ with
$I_{\mathrm{par}}=10.11\%$---the highest among all variants in Fig.~\ref{fig:model_comparison}---indicating that
the shell and hexadecapole effects are nearly additive and largely independent.
Adding $\beta_2^{(p)}$ or $\beta_4^{(d)}$ on top of this produces only marginal
further gains while degrading $I_{\mathrm{par}}$.  Equation~(\ref{eq:improved})
therefore offers the optimal balance of accuracy, parsimony, and physical
interpretability.

Nevertheless, the two corrections are not fully decoupled in the fit.  The
coefficient $d$ changes from $-6.712$ (without $C_{pn}$) to $-5.986$ (with
$C_{pn}$), a reduction of about $11\%$ in magnitude.  This shows that, in the
absence of an explicit shell correction, part of the shell-dependent residual is
absorbed into the deformation term.  For a typical
$\beta_4^{(p)}\approx0.05$--$0.1$, the spurious shell contribution amounts to
$0.04$--$0.07$ in $\log_{10}T_{1/2}$.  Crucially, the sign and the order of
magnitude of $d$ remain stable, so the hexadecapole dependence is not an
artifact of the missing shell term.

\begin{figure}[htp]
	\centering
	\includegraphics[width=1\linewidth]{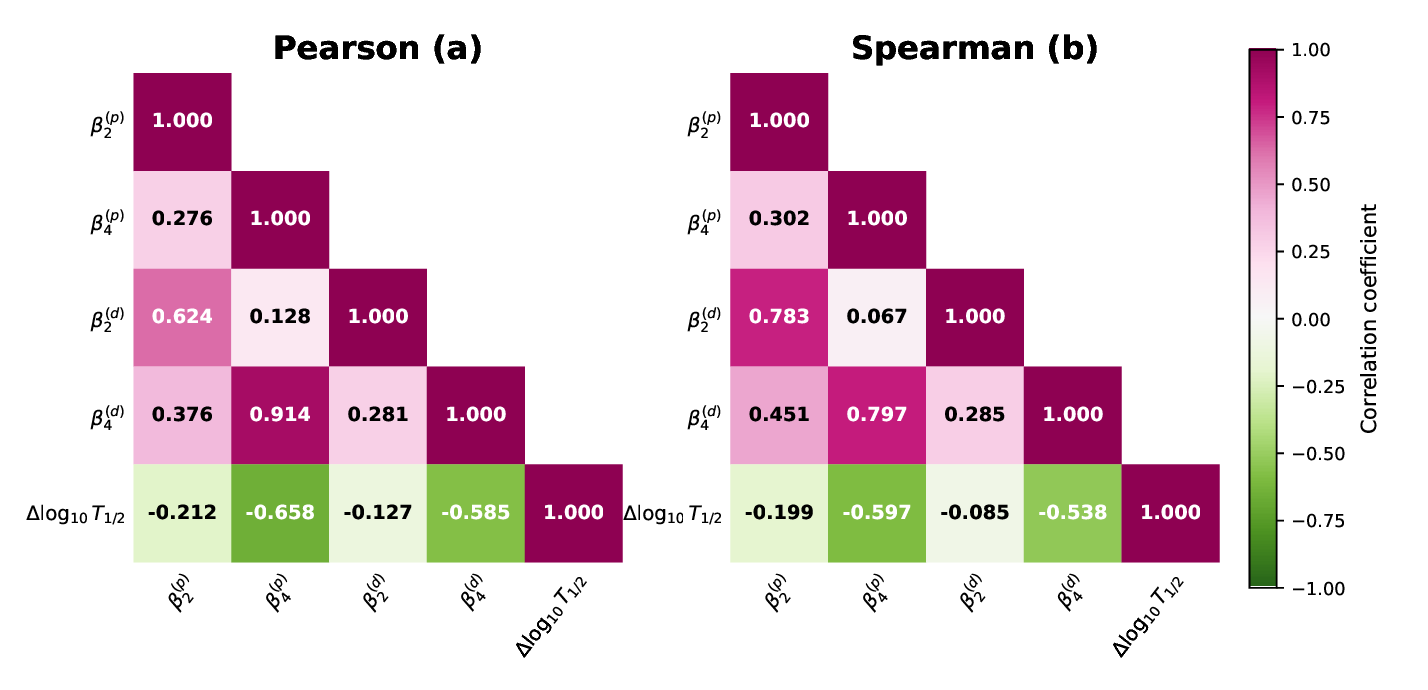}
	\caption{(Color online) Pearson (a) and Spearman (b) correlation matrices among the parent-
	and daughter-nucleus deformation parameters $\beta_2^{(p)}$, $\beta_4^{(p)}$,
	$\beta_2^{(d)}$, $\beta_4^{(d)}$, and the residual $\Delta\log_{10}T_{1/2}$
	of Eq.~(\ref{eq:residual}).}
	\label{fig:correlation}
\end{figure}

To identify which deformation parameter carries the remaining structural
information, we remove the dominant shell trend by defining
\begin{equation}
\Delta\log_{10}T_{1/2}
= \log_{10}T_{1/2}^{\rm exp} - \log_{10}T_{1/2}^{(0)},
\label{eq:residual}
\end{equation}
where $T_{1/2}^{(0)}$ is obtained from the Royer${}+C_{pn}$ baseline.  The
correlation matrices are shown in Fig.~\ref{fig:correlation}.  Among the four
deformation parameters, $\beta_4^{(p)}$ correlates most strongly with the
residual (Pearson $-0.658$, Spearman $-0.597$), far exceeding
$\beta_2^{(p)}$ ($-0.212$, $-0.199$) and $\beta_2^{(d)}$ ($-0.127$,
$-0.085$).  The correlation of $\beta_4^{(d)}$ ($-0.585$, $-0.538$) is
comparable but not superior; moreover, $\beta_4^{(d)}$ is itself strongly
correlated with $\beta_4^{(p)}$ ($0.914$, $0.797$), reflecting the shape
inheritance between parent and daughter.  The two choices are therefore close to
redundant, and $\beta_4^{(p)}$ is preferred both for its stronger correlation and
for its more direct physical connection to the preformation process.

\subsection{Complementary roles of shell and deformation effects}
\begin{figure}[htp]
	\centering
	\includegraphics[width=1\linewidth]{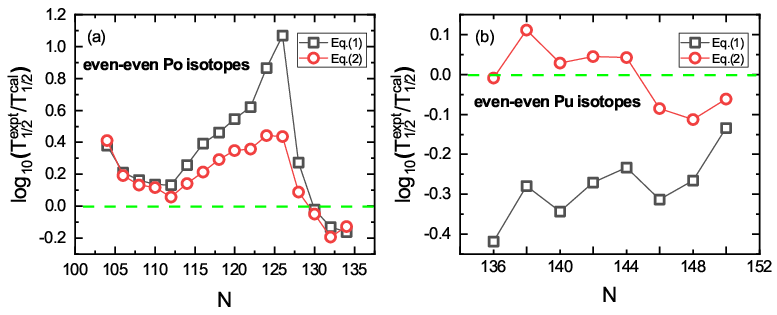}
	\caption{(Color online) Logarithmic deviations between the experimental and calculated
	$\alpha$-decay half-lives as functions of the neutron number $N$ for
	(a) even--even Po isotopes and (b) even--even Pu isotopes. Black open squares
	and red open circles denote the results of Eq.~(\ref{eq:royer}) and
	Eq.~(\ref{eq:improved}), respectively; the green dashed line marks zero
	deviation.}
	\label{fig:isotopes}
\end{figure}

The two terms act in distinct regions of the nuclear chart, as illustrated in
Fig.~\ref{fig:isotopes}.  For the Po isotopes, which straddle $N=126$, the
Royer residuals peak sharply at the shell closure, and
Eq.~(\ref{eq:improved}) removes most of this peak, confirming that $C_{pn}$
corrects the shell effect.  For the Pu isotopes, far from any magic number,
$C_{pn}$ brings little improvement and the reduction comes almost entirely from
$\beta_4^{(p)}$. In short, $C_{pn}$ dominates near shell
closures and $\beta_4^{(p)}$ in open-shell regions, which is why neither term
alone suffices over the whole data set.

\begin{figure}[htp]
	\centering
	\includegraphics[width=1\linewidth]{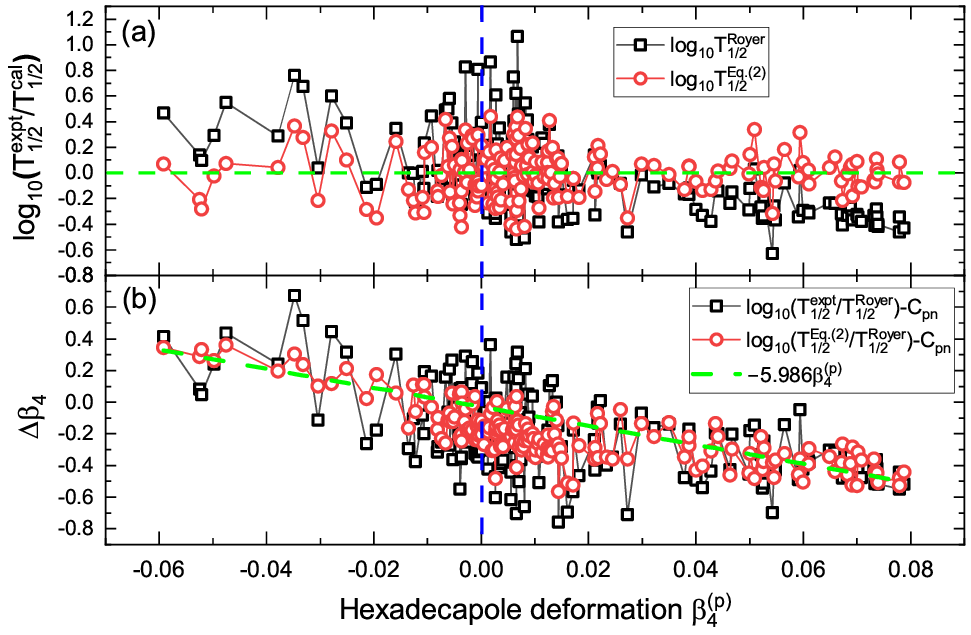}
	\caption{(Color online) (a) Logarithmic half-life residuals as functions of the parent-nucleus
	hexadecapole deformation $\beta_4^{(p)}$. Open squares and open circles denote Eqs.~(\ref{eq:royer}) and (\ref{eq:improved}). Dashed vertical lines indicate \(\beta_{4}^{(p)}=0\) and the zero‑residual reference.
  (b) Shows the same quantities after subtracting $C_{pn}$. Green dashed line is the linear fitting of the red circle.}
	\label{fig:192}
\end{figure}

The sign dependence of the hexadecapole correction is displayed in
Fig.~\ref{fig:192} for the full data set.  Panel~(a) shows that
Eq.~(\ref{eq:improved}) compresses the residuals toward zero over the entire
range of $\beta_4^{(p)}$.  To separate this from the dominant shell
contribution, panel~(b) shows the same quantities with $C_{pn}$ subtracted: a
clear decreasing trend with $\beta_4^{(p)}$ survives, in accordance with the
negative fitted coefficient $d=-5.986$.  The persistence of this trend demonstrates that
$\beta_4^{(p)}$ carries structural information genuinely distinct from the shell
effect: positive $\beta_4^{(p)}$ shortens half-lives, as expected if locally
convex surface regions (cf.~Fig.~\ref{fig:shape}) favor $\alpha$-cluster
preformation, while negative $\beta_4^{(p)}$ acts in the opposite direction.
This interpretation is supported by machine-learning
analyses~\cite{Ma2026}, which report a strong correlation between hexadecapole
deformation and the $\alpha$-preformation probability.

\subsection{Robustness}
\begin{table}[htbp]
	\caption{Fitted parameters and RMSD of Eq.~(\ref{eq:improved}) obtained with
	the parent-nucleus hexadecapole deformation $\beta_4^{(p)}$ from different
	mass models.}
	\label{tab:models}
	\begin{ruledtabular}
		\begin{tabular}{lccccc}
			Model & $a$ & $b$ & $c$ & $d$ & RMSD \\
			\hline
			WS4~\cite{Wang2014}      & $-26.718\pm0.151$ & $-1.138\pm0.006$ & $1.608\pm0.003$ & $-5.986\pm0.476$ & $0.184$ \\
			WS3.3~\cite{Wang2010WS3}    & $-26.693\pm0.153$ & $-1.139\pm0.006$ & $1.607\pm0.004$ & $-5.488\pm0.454$ & $0.187$ \\
			KTUY~\cite{Koura2005}     & $-26.475\pm0.162$ & $-1.148\pm0.006$ & $1.607\pm0.004$ & $-5.323\pm0.523$ & $0.200$ \\
			FRDM2012~\cite{Moller2016} & $-26.605\pm0.151$ & $-1.145\pm0.006$ & $1.609\pm0.004$ & $-3.816\pm0.308$ & $0.185$ \\
		\end{tabular}
	\end{ruledtabular}
\end{table}
Since $\beta_4^{(p)}$ is not a directly measured quantity, we repeated the fit
using values from the WS4~\cite{Wang2014}, WS3.3~\cite{Wang2010WS3},
KTUY~\cite{Koura2005}, and FRDM2012~\cite{Moller2016} mass models
(Table~\ref{tab:models}).  The RMSD lies between $0.184$ and $0.200$ in all
four cases, and $d$ remains negative throughout; neither the quality of the
description nor the sign of the effect depends on the choice of mass model.  The
variation in the magnitude of $d$ (from $-5.986$ for WS4 to $-3.816$ for
FRDM2012) reflects the different $\beta_4$ scales of the underlying models
rather than a different physical trend.

The correction is also not tied to the Royer formula.
Table~\ref{tab:formula_comparison} lists the RMSDs obtained when
$d\beta_4^{(p)}+C_{pn}$ is appended to the
Viola--Seaborg--Sobiczewski~(VSS)~\cite{Viola1966}, NRDX~\cite{Ni2008},
universal decay law (UDL)~\cite{Qi2009}, and UDL-FF~\cite{Ismail2025} formulas,
with all coefficients refitted to the same 192 even--even nuclei.  The reductions
range from $21.78\%$ to $42.90\%$.  The UDL-FF case is particularly
instructive: replacing its daughter-nucleus deformation terms
$\beta_2^{(d)}+\beta_4^{(d)}$ by $d\beta_4^{(p)}+C_{pn}$ lowers the RMSD from
$0.225$ to $0.176$ while reducing the number of deformation-related
coefficients from two to one.

\begin{table}[htbp]
	\caption{RMSDs of different empirical formulas before and after including the
	shell--deformation correction $d\beta_4^{(p)}+C_{pn}$. All coefficients are
	refitted to the same set of 192 even--even nuclei. The improvement denotes
	the relative reduction in RMSD.}
	\label{tab:formula_comparison}
	\begin{ruledtabular}
		\begin{tabular}{lccc}
			Formula & Original RMSD & Corrected RMSD & Improvement (\%) \\
			\hline
			Royer~\cite{Royer2000}  & 0.309 & 0.184 & 40.45 \\
			UDL~\cite{Qi2009}        & 0.317 & 0.181 & 42.90 \\
			VSS~\cite{Viola1966}     & 0.344 & 0.212 & 38.37 \\
			NRDX~\cite{Ni2008}       & 0.368 & 0.231 & 37.23 \\
						UDL-FF~\cite{Ismail2025} & 0.225 & 0.176 & 21.78 \\
		\end{tabular}
	\end{ruledtabular}
\end{table}
These results resolve the apparent conflict with Ismail \textit{et
al.}~\cite{Ismail2025}, who concluded that $\beta_4$ has only a marginal effect.
Two key differences account for this discrepancy.  First, they employed the
daughter deformation, which our correlation analysis shows to be the less
effective choice.  Second, their formulas contain no explicit shell correction,
so part of the shell residual is absorbed into the deformation coefficients,
obscuring the genuine $\beta_4$ dependence. Once the shell effect is isolated
by $C_{pn}$, the hexadecapole contribution emerges clearly: the fitting,
correlation, and isotopic-chain analyses all converge on the same picture, in
which $C_{pn}$ removes the dominant shell trend and $\beta_4^{(p)}$ accounts
for the remaining structure-dependent residuals, together yielding a significant
and physically interpretable improvement in $\alpha$-decay systematics.

\subsection{Predictions for superheavy nuclei}
\begin{figure}[htp]
	\centering
	\includegraphics[width=1\linewidth]{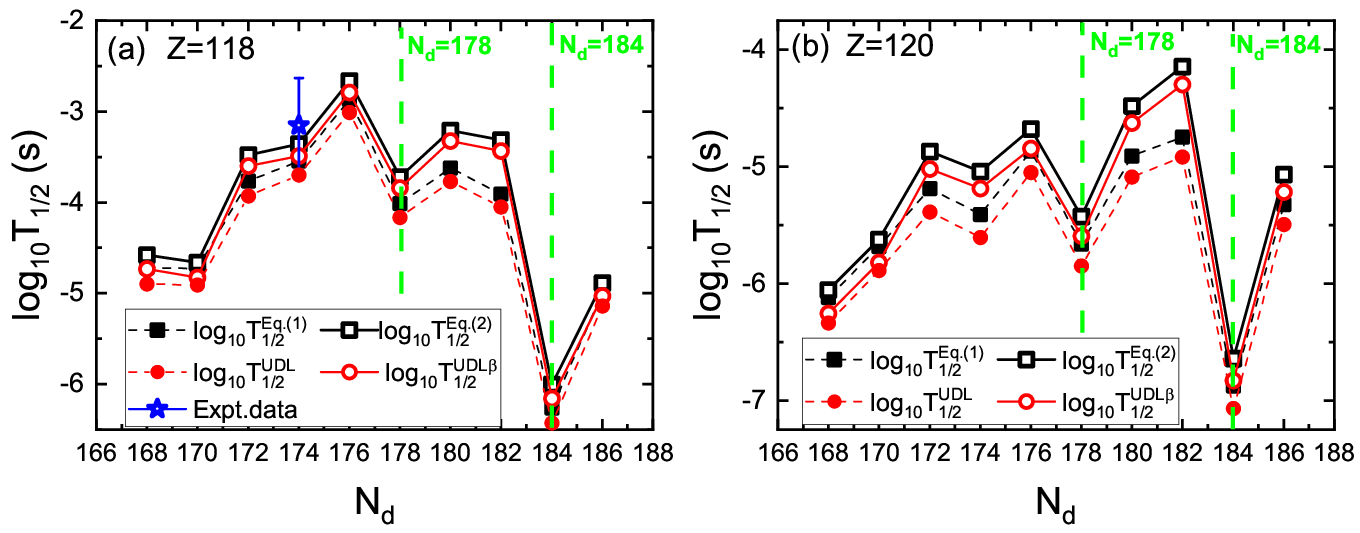}
	\caption{$\log_{10}T_{1/2}$ of the $Z=118$ (a) and $Z=120$ (b) isotopes as
	functions of the daughter neutron number $N_d$. Predictions of
	Eq.~(\ref{eq:improved}) (open squares) are compared with the original Royer
	formula, Eq.~(\ref{eq:royer})~\cite{Royer2000} (solid squares), the UDL,
	Eq.~(\ref{eq:UDL})~\cite{Qi2009} (solid circles), and the UDL$\beta$,
	Eq.~(\ref{eq:UDLbeta}) (open circles). The experimental value for
	$^{294}$Og (open star) is taken from NUBASE2020~\cite{nubase2020}. Vertical
	dashed lines mark the predicted shell closures at $N_d=178$ and	$N_d=184$.}
	\label{fig:yuce}
\end{figure}
An improved empirical formula is ultimately valuable only if it yields testable
predictions for superheavy nuclei, where experimental data remain scarce.
Figure~\ref{fig:yuce} shows the predicted $\alpha$-decay half-lives of thee
$Z=118$ (a) and $Z=120$ (b) isotopic chains as functions of the edaughter
neutron number $N_d$, obtained from Eqs.~(\ref{eq:royer}),
(\ref{eq:improved}), (\ref{eq:UDL}), and (\ref{eq:UDLbeta}).  For nuclei
without measured $Q_\alpha$ values, the decay energies are taken from the
WS4+AKRR mass model~\cite{Tian25}. The only known datum in these chains, $^{294}$Og ($N_d=174$),
provides a direct benchmark: the original Royer and UDL formulas underestimate
its half-life, whereas Eq.~(\ref{eq:improved}) and the UDL$\beta$ formula both
reproduce the measured value. That two independently corrected formula agree
with this datum and with each other indicates the improvement stems from the
correction term $d\,\beta_4^{(p)}+C_{pn}$ itself rather than from a particular
parameterization.

The predicted trends exhibit clear shell-structure signatures, with pronounced
minima around $N_d=178$ (deformed sub-shell closure) and prominent features at
$N_d=184$ (spherical neutron shell closure). The corrected formulas sharpen
these features relative to the uncorrected ones, as expected from the explicit
$C_{pn}$ term. For the yet-to-be-synthesized $Z=120$ isotopes, the two
corrected formulas give mutually consistent predictions that smoothly extend
the $Z=118$ systematics, awaiting confrontation with future synthesis
experiments.
\section{Summary}
We show that positive parent-nucleus hexadecapole deformation
($\beta_4^{(p)}>0$) systematically enhances $\alpha$ decay by favoring
cluster preformation at locally convex surface regions. This conclusion
requires proper separation of shell and deformation effects: without
$C_{pn}$, shell-induced residuals near magic numbers are absorbed into
the deformation term, obscuring the genuine $\beta_4$ dependence. Once
$C_{pn}$ removes the dominant shell trend, $\beta_4^{(p)}$ emerges as
the deformation parameter most strongly correlated with the remaining
residuals, outperforming $\beta_2^{(p)}$ and daughter-nucleus
deformations---consistent with the fact that the $A^{1/6}\sqrt{Z}$ term
already absorbs the dominant quadrupole effect while $\beta_4^{(p)}$
captures local surface curvature inaccessible to that smooth term. The
resulting sign-dependent correction reduces the RMSD from $0.309$ to
$0.184$ for 192 even--even nuclei with only four adjustable parameters.
In the superheavy region, the formula reproduces the measured half-life
of $^{294}$Og and yields consistent predictions for the $Z=120$
isotopic chain. Predicted half-lives for 1060 even--even nuclei are
provided as Supplemental Material~\cite{Tianjl26}.

\begin{center}
\textbf{ACKNOWLEDGMENTS}
\end{center}
This work was supported by the Guangxi Science and Technology Program (No. 2023GXNSFDA026005 and No. 2023GXNSFBA026008), the National Natural Science Foundation of China (No. 12465019 and No. 12465021),  the Central Government Guides Local Scientific and Technological Development Fund Projects (No. Guike ZY22096024), and the Innovation Project of Guangxi Graduate Education (No. YCBZ2026099)

\end{document}